# Accelerated Aging in 3-mol%-Yttria-Stabilized Tetragonal Zirconia Ceramics Sintered in Reducing Conditions

José F. Bartolomé [a], Isabel Montero [a], Marcos Díaz [a], Sonia López-Esteban [a], José S. Moya [a], Sylvain Deville [b], Laurent Gremillard [b], Jérôme Chevalier [b], Gilbert Fantozzi [b]

[a] Instituto de Ciencia de Materiales de Madrid (ICMM), Consejo Superior de Investigaciones Cientificas (CSIC), Cantoblanco 28049, Madrid, Spain
[b] Institut National des Sciences Appliquées (INSA), Groupe d'Etudes de Métallurgie Physique et de Physique des Matériaux (GEMPPM), UMR 5510, 69621 Villeurbanne, France

## Abstract

The aging behavior of 3-mol%-yttria-stabilized tetragonal zirconia (3Y-TZP) ceramics sintered in air and in reducing conditions was investigated at 140°C in water vapor. It was observed by X-ray diffraction (XRD) that 3Y-TZP samples sintered in reducing conditions exhibited significantly higher tetragonal-to-monoclinic transformation than samples with similar density and average grain size values but obtained by sintering in air. This fact is explained by the increase of the oxygen vacancy concentration and by the presence at the grain boundary region of a new aggregate phase formed because of the exolution of $Fe^{2+}$ ions observed by X-ray photoelectron spectroscopy.

The aging degradation of yttria-stabilized tetragonal zirconia ceramics (Y-TZP) is considered today an issue of important technological interest, mainly if long-term performance is required,[1] e.g., hip and knee prostheses (>10 years) or solid oxide fuel cells (SOFCs).

Recently, alarming problems related to the aging of the 3Y-TZP (zirconia doped with 3 mol% $Y_2O_3$) femoral head in total hip replacement have been reported.[2] In particular, resistance to steam sterilization and the hydrothermal stability of yttria-containing zirconia in the body have been questioned. Aging occurs by a tetragonal-to-monoclinic (t–m) phase transformation of grains on any surface in contact with water or body fluids which involves surface roughening, grain pull-out, and the formation of microcracks on the specimen surface and consequently strength degradation. This accelerated low-temperature degradation (LTD) leads to premature failure of components.

The phase stability of YSZ ceramic electrolytes is of key importance to long-term SOFC applications. Current efforts are aimed at lowering the operating temperature of solid oxide fuel cells from above 900° to 500°C to improve the longevity and cost. of the peripheral materials and the electrical power generation efficiency.[3] Tetragonal zirconia is considered a promising candidate to be used as electrolyte in intermediate-temperature solid oxide fuel cells (IT-SOFC), because of its good mechanical and electrical properties in comparison with cubic zirconia at lower temperatures.[4] However, the low-temperature degradation caused by tetragonal-to-monoclinic transformation under hydrothermal conditions leads to physical degradation; this makes its application in IT-SOFC devices challenging as water vapor is produced at the anode when the cell is in operation. Moreover, the fuel cell must be able to withstand thermal cycling and the ability to operate under pressures higher than atmospheric pressure over the lifetime of the cell (>50 000 h).[5] Therefore, the fuel-cell components are susceptible to slow crack growth (SCG).

Considerable efforts have been exerted to elucidate the mechanism of hydrothermal degradation in Y-TZP. The yttrium depletion by $Y(OH)_3$ formation in the presence of water vapor was proposed by Lange et al.[6] as the cause of the aging-induced degradation. However, this mechanism cannot explain the effect of grain size.[7] Moreover, it has been demonstrated that the degradation occurs even in dry air and under vacuum, depending on the type and content of alloying elements in Y-TZP, and the loss of $Y_2O_3$ is not involved in LTD when aged in air.[8] Sato et al.[9] suggested that the hydrothermal degradation of Y-TZP is controlled by the breakage of Zr–O–Zr bonds to form Zr–OH due to the reaction between the chemisorbed $H_2O$ and $Zr^{4+}$ at the specimen surface, which results in the release of strain energy that would ensue if the t–m transformation would occur. However, this model cannot rationalize the composition and grain-size dependence of the degradation rate. Yoshimura[10] proposed that the accumulated strain area resulting form the migration of OH$^-$ at the surface and in the lattice serve as a nucleus of the m-ZrO$_2$ phases in the t-ZrO$_2$ matrix. However, studies on oxygen diffusion in Y-TZP and the effect of oxygen on the tetragonal-to-monoclinic transformation pose problems for the Yoshimura et al. theory.[11,12] Recently, Kim et al.[8] reported that LTD of Y-TZP is governed not by the existence of H$_2$O but the relaxation process of internally strained lattice due to a thermally activated oxygen vacancy diffusion from the surface into the interior of the specimen. There is a general agreement in literature that the degradation is accelerated as the material is exposed to water or water vapor because the residual stress facilitates the t–m phase transformation and the reaction between the Zr–O–Zr bond and H$_2$O. Therefore, the rate of LTD is governed by both the number of oxygen vacancies and the instability of t-ZrO$_2$.

It has also been pointed out that modifying the grain boundaries by doping or by an inhomogeneous yttria concentration [13] is thought to play an important role in governing the hydrothermal degradation of Y-TZP ceramics.

In the SOFCs, nickel (Ni) and Y$_2$O$_3$ -stabilized zirconia (YSZ) cermet is widely used as an anode material because of its high catalytic property and low cost.[14] A general method currently used for anode manufacturing is mechanical mixing of NiO and Y$_2$O$_3$–ZrO$_2$ powders, pasting mixing composite powders on a zirconia electrolyte plate by screen printing or spraying and then sintering under reducing atmosphere to form the Ni/YSZ cermets.[15] The phenomenon of aging-induced tetragonal-to monoclinic phase transformations in zirconia ceramics sintered in air has been widely documented; however, a direct study of the accelerated aging process in yttria-stabilized tetragonal zirconia ceramics sintered under reducing conditions is still lacking. The aim of the present work is to explore the effect of the sintering atmosphere on the accelerating aging of yttria-stabilized tetragonal zirconia ceramics.

## II. Experimental Procedure

The following commercially available tetragonal zirconia polycrystals powder (Y-TZP 3 mol.%; TZ-3YS, Tosoh Corp.) has been used as starting material with an average particle size of d$_{50}$ =0.2 µm, a BET specific surface area of 6.7 m$^2$/g, and the following chemical analysis (wt%): ZrO$_2$ (95.00), Y$_2$O$_3$ (4.98), Al$_2$O$_3$ (0.005), Fe$_2$O$_3$ (0.005), SiO$_2$ (0.002), and Na$_2$O (0.003).

The powders were isostatically pressed at 200 MPa. The resulting cylindrical bars (6-mm diameter and 50-mm length) were fired in a 90% Ar/10% H$_2$ atmosphere or in air at 1430°C for 2 h. The heating and cooling rate was kept at 600°C/h. The bulk densities of all the samples were measured using the Archimedes method.

SEM micrographs were obtained using a HITACHI S-4700 field emission scanning electron microscope (Japan) with an accelerating tension of 20 kV. Polished samples were preliminarily thermally etched in a 90% Ar/10% $H_2$ or in air atmosphere at 1400°C for 30 min, with heating and cooling rates of 300°C/h. Samples were then gold-coated before SEM analysis. The grain size of the 3Y-TZP was determined using a linear-intercept method on representative SEM micrographs.

X-ray photoelectron spectra (XPS) were recorded with a hemispherical electron analyzer (Leybold 10) equipment. The measurements were done with $MgK_\alpha$ excitation (hv=253.6 eV) in vacuo at <$10^{-9}$ Pa at room temperature. The acceleration voltage and current of the non monochromatized X-ray source were 11 kV and 20 mA, respectively. The samples showed charging of <2eV in XP spectra because of their insulator nature. The binding energies were corrected for charging effect by assuming a constant binding energy of the C 1s peak position of graphitic carbon at 284.6 eV.

Aging experiments were conducted using a steam autoclave at 140°C, with a 2-bar pressure. Polished samples were located in steam autoclaves (Fisher Bioblock Scientific, France) and left in steam atmosphere for different periods (up to 20 h). Samples were located on a grid in the autoclave so that they were not soaked in water during aging, but only subjected to steam atmosphere. Before increasing the temperature to 140°C, the enclosure was left open to evacuate the air atmosphere initially present in the steam autoclave, thus ensuring a 100% humidity atmosphere during aging.

X-ray diffraction data were obtained with a diffractometer using Ni-filtered $CuK_\alpha$ radiation. The tetragonal/monoclinic zirconia ratio was determined using the integrated intensity (measuring the area under the diffractometer peaks) of the tetragonal (101) and two monoclinic (111) and (1 11) peaks as described by Toraya et al.[16] Diffractometer scans were obtained from 27° to 33°, at a scan speed of 0.2°/min and a step size of 0.02°.

## III. Results and Discussion

The density of the sintered 3Y-TZP samples was found to be approx. 98%. Figure 1 shows SEM micrographs of the thermally etched polished surfaces of both a $ZrO_2$ sample sintered in air and in 90%Ar/10% $H_2$ atmosphere. In both cases the average grain size of zirconia was found to be about 0.3 µm.

The monoclinic phase volume fraction evolution during aging at 140°C in steam for the two ceramics is shown in Fig. 2. For both 3Y-TZP samples, the starting surfaces after polishing are free of monoclinic phase. The m-$ZrO_2$ fraction increases at a quite-steady rate with aging time, reaching 35% for the sample sintered in air and 55% for the one sintered in 90%Ar/10%H$_2$ atmosphere after 20 h of aging at 140°C.

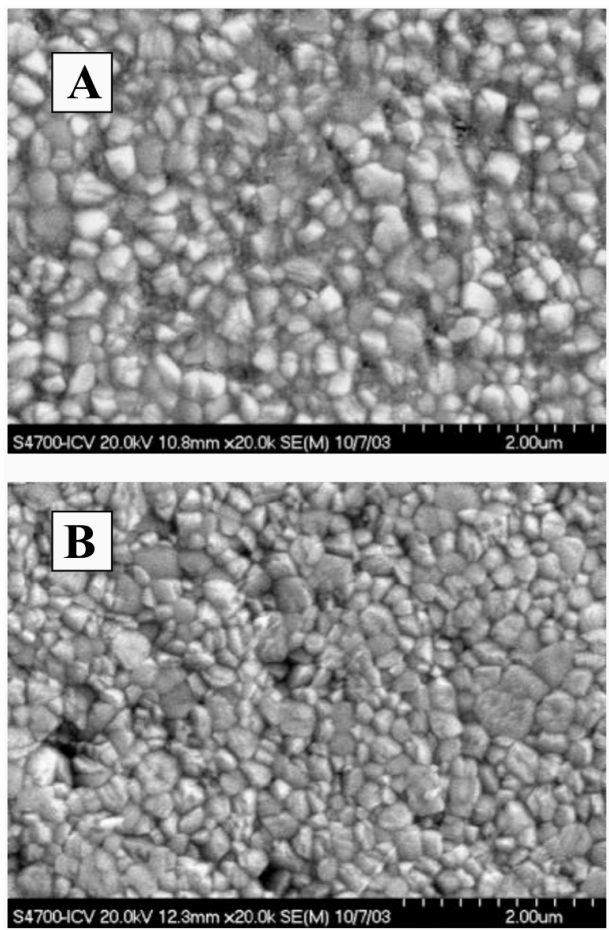

Fig. 1. Polished surfaces of (A) the ZrO$_2$ sample sintered in air and (B) the ZrO$_2$ sample sintered in 90%Ar/10% H$_2$ atmosphere, after being thermally etched at 1400°C for 0.5 h in air and reducing conditions, respectively.

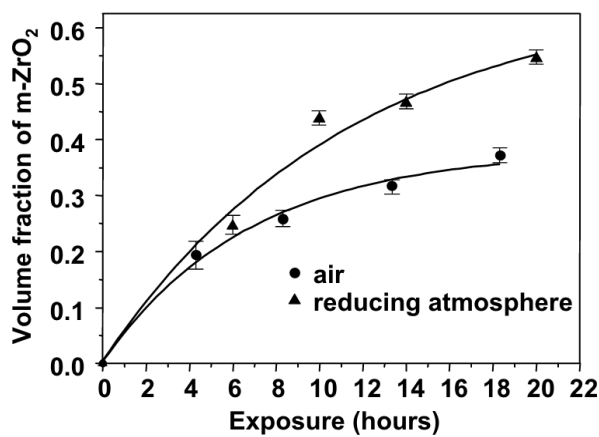

Fig. 2. Surface monoclinic phase fraction versus exposure time at 140°C in steam for the zirconia materials sintered in air and in 90%Ar/10% H$_2$ atmosphere.

It is known that the oxygen vacancies created either by reducing the oxygen partial pressure or by alloying zirconia with aliovalent metal oxides ($Mg^{2+}$, $Ca^{2+}$, and $Y^{3+}$) play an important role in stabilizing the zirconia structure.[17] Therefore, in our 3Y-TZP samples, in addition to the oxygen vacancies produced by substitution of yttrium cations for zirconium cations to maintain electrical neutrality, which are expected to have a negative charge, uncharged oxygen vacancies are introduced by sintering in a low oxygen partial pressure in response to thermodynamic considerations. The concentration of thermodynamic oxygen vacancies appears to be linked to oxygen partial pressure with the following empirical semi-quantitative relationship:

$$[V_o^{''}] \propto [O_2]^{-1/6}$$

where $[V_o^{''}]$ and $[O_2]$ represent the concentration of thermodynamic oxygen vacancies and the oxygen partial pressure, respectively. We can roughly calculate the concentration of thermodynamic oxygen vacancies.[19] A pressure of about $10^{-24}$ Pa at 1400°C, which can be readily achieved in a hydrogen atmosphere, leads to about 2000 ppm intrinsic thermodynamic oxygen vacancies. Then, in the sample obtained under reducing conditions, the concentration of oxygen vacancy created by the $Y^{3+}$ doping (-15 000 ppm) increases by the sintering in reducing atmosphere, leading to a high vacancy diffusion rate which governed the t–m phase transformation during aging at low temperatures. This statement is in agreement with the X-ray irradiation experiment in air performed by Gremillard[20] on the 3Y-TZP sample surface. After 1h in an autoclave at 134°C in water vapor, he observed that the color changes from dark to white due to the elimination of oxygen vacancies.

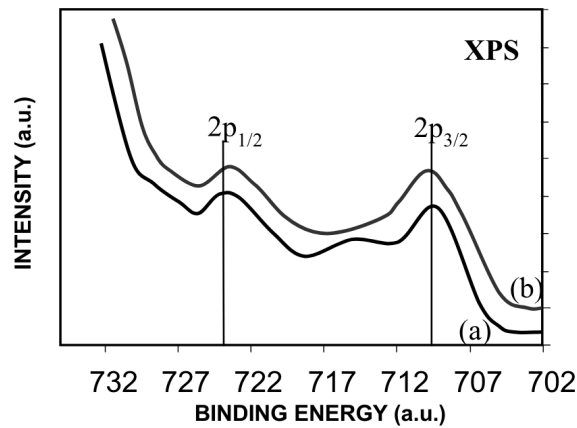

Fig. 3. XPS of the Fe 2p core level for the zirconia samples obtained after air and $H_2$ treatments, curves a and b, respectively. Spectra are displaced in intensity scale to allow better observation.

On the other hand, if there was significant change in grain boundary composition, it can be inferred that the grain boundary regions are active sites for monoclinic nucleation during aging. The near-surface region of the ceramics treated in 90% Ar/10% $H_2$ atmosphere and in air was analyzed by XPS. Figure 3 shows the Fe 2p core level spectra of the ceramics treated in 90% Ar/10% $H_2$ atmosphere and in air, curves a and b, respectively. The position of the Fe 2p 3/2 peak at a binding energy of about 710.5 eV is characteristic of the iron oxide.[21] Both peaks Fe $2p_{3/2}$ and $2p_{1/2}$ are accompanied by satellites structures on their high binding energy side. These satellite structures in the XPS spectra were caused by the Fe 3d–O 2p hybridization. In the case of the spectrum a, the satellite of the $2p_{3/2}$ main peak is clearly observed on their high binding energy side, at about 7 eV. This result indicates the presence of $Fe^{3+}$.[22] In contrast, the signal of the $Fe^{3+}$ satellite in spectrum b has smeared out. It can be due to the major contribution of the $Fe^{2+}$ satellite at the spectrum. Thus, the shoulder of the Fe $2p_{3/2}$ toward higher energies (spectrum b) is characteristic of the formation of $Fe^{2+}$ ions. In addition, the Fe $2p_{1/2}$ peak position shifts toward low binding energies about 1 eV.[22] Moya et al.,[23] using electron spin resonance techniques, detected the exolution of iron impurities (ppm) which are present at the starting powders, from the bulk to the grain boundary of yttria-partially-stabilized zirconia (3Y-TZP) polycrystals under reducing conditions. The isolated $Fe^{3+}$ ions are removed from the $ZrO_2$ lattice, by the reduction, forming a new aggregate phase at the grain boundaries in the polycrystalline zirconia. This process can be understood by considering that reduction of $Fe^{3+}$ to $Fe^{2+}$ increases ion size and consequently diminishes the solid solution of iron. The presence of $Fe^{2+}$ in the sample sintered in reducing atmosphere has been clearly established by XPS (Fig. 3). The segregation of $Fe^{2+}$ ions at the grain boundaries and their subsequent aggregation in a separate phase, i.e., magnetite, may produce a volume expansion, as has been reported by Chein and Ko [24] On the other hand, the formation of a second phase in the grain boundary may result in a decrease of the surface energy. A monoclinic phase at room temperature is then formed when this free-energy term is low enough.[25] Therefore in the sample sintered in reducing atmosphere the aging process can be accelerated by both cooperating mechanisms, increasing the bulk concentrations of oxygen vacancies and at the same time modifying the grain boundary region by the exolution of $Fe^{2+}$ ions.

## IV. Conclusions

It has been observed an aging rate increase of about 50% in a 3Y-TZP dense sample sintered in reducing conditions versus a sample with similar density and average grain size values but obtained by sintering in air. This fact is explained in terms of the following cooperative mechanisms: (a) the increase of the oxygen vacancy concentration that is introduced by sintering in a low oxygen partial pressure, leading a high vacancy diffusion rate which governed the t–m phase transformation during aging at low temperatures, and (b) the presence at the grain boundaries in the polycrystalline zirconia of a new aggregate phase, i.e., magnetite, formed because of the exolution of $Fe^{2+}$ ions that promote a volume expansion affecting the stability of the tetragonal phase.

Technology of Zirconia III. Edited S. Somiya, N. Yamamoto, and H. Yanagida. American Ceramic Society, Westerville, OH, 1988.